# On the Role of Identity in Surveillance

Victoria Wang & John V. Tucker

Abstract

Surveillance is a process that observes behaviour, recognises properties and identifies individuals. It has become a commonplace phenomenon in our everyday life. Many surveillance practices depend on the use of advanced technologies to collect, store and process data. We propose (i) an abstract definition of surveillance; and (ii) an abstract definition of identity, designed to capture the common structure of many disparate surveillance situations. We argue that the notion of identity is fundamental to surveillance. Rather than having a single identity, individuals have many identities, real and virtual, that are used in different aspects of their lives. Most aspects of life are subject to some form of surveillance, and observations and identities can be aggregated. The notion of identity needs to be theorised. Our analysis is very general and, at the same time, sufficiently precise to be the basis of mathematical models.

**Key Words**: Surveillance, Identity, Social Sorting, Technology, Security

## 1. Introduction

Once, the word surveillance was reserved for criminological contexts, such as the traditional scrutiny of suspects or contemporary pre-crime practices. Nowadays, the notion is much more general as surveillance practices are increasingly becoming a natural component of our everyday life. Modern marketing and customer relations management depend upon the effective monitoring and classification of individuals' interests and tastes. Such surveillance can be of a high standard and can certainly expedite their everyday lives. The classification of individuals into categories is a necessary consequence, explicit or implicit, of surveillance. In pre-crime, the categories are used for the assignment of risk. In all cases, the categories affect individuals' opportunities or treatments.

David Lyon has emphasised a general conception of surveillance, which he has characterised as "the focused, systematic and routine attention to personal details for purposes of influence, management, protection or detection" (2007a: 14). Furthermore, "this attention to personal details is not random, occasional or spontaneous; it is deliberate and depends on certain protocol and techniques" (ibid.). Lyon (2003, 2007b) has emphasised the significance of considering contemporary surveillance as *social sorting*. He defined the term to mean the "focus on the social and economic categories and the computer codes by which personal data is organized with a view to influencing and managing people and populations" (Lyon, 2003: 2). Social sorting has become the main purpose of surveillance, since surveillance today is overwhelmingly about personal data.

The rise of surveillance leads to an emphasis on monitoring the behaviours from selected individuals, through groups of people, to the entire population. The growth and effectiveness of the monitoring are made possible by all sorts of new technologies, especially software technologies. However, surveillance as social sorting is becoming increasingly significant, not merely because of the abundance and availabilities of new technological devices. Rather, these devices are required because of the increasing number of perceived and actual risks, and consequently, the desire to monitor the behaviour of the entire population (cf. Lyon, 2003).

The notion of identity is to be found at the heart of all contemporary surveillance practices (Ball et al., 2012). Many forms of surveillance systems are supported by identity management systems that depend on technologies to provide information on identity. The systems of fingerprinting (Cole, 2001), iris scans (Lyon, 2001), facial recognition systems (Introna & Wood, 2004), DNA samples (Wallace, 2006), passport (Torpey, 2000) and National Identity Register (NIR) (UK Government, 2006) are used for accuracy of identification. The UK Government's Foresight Programme investigated the future of identity in a major project in order to explore how transformations in technology, politics, economics, our environment and demographics will change our notion of identity (Foresight, 2013). The rapid speed of developments in technology, especially those brought about the Internet, has been identified as the key driver in the transformation, and thus, the problematization of identity. Notions of identity are central to surveillance, and nowadays, the notions include both data from embodied individuals, and increasingly, data about these individuals to be found circulating in the Internet (Lyon, 2003). Actually, due to the increasing frequency of non-face-to-face interactions, the Internet has become the main method of categorising.

Notions of identity are essential for surveillance practices, since social sorting – the main purpose of contemporary surveillance – is about the classification and categorisation of personal data. Sociologies of surveillance as social sorting seem to remain underdeveloped (Lyon, 2003). In his article 'Surveillance, Security and Social Sorting: Emerging Research Priorities', Lyon calls for multidisciplinary approaches towards surveillance to achieve a better understanding of, amongst other things, "the dynamics of social sorting and the social implications of the new technologies" (2007b: 168). This article offers a theoretical analysis of surveillance focusing on identity. Although tailored to the use of software technologies, the analysis covers a wide range of surveillance practices. Its purpose is to find a concise and compact set of concepts that can be used to model most forms of surveillance. Specifically, we formulate:

(i) an abstract definition of surveillance that works in both the physical and virtual worlds; and
(ii) an abstract definition of identity that applies to people and objects.

The definitions are intended to capture and structure the essential ideas that are common to many disparate situations of surveillance. We argue that the notion of identity is fundamental to surveillance. We propose that surveillance and identity are fundamentally a matter of assigning data to people and objects:

(a) the technologies for observing people and objects capture behaviour in data; and

(b) the identity of objects and people are defined by means of data which we call *identifiers*.

Our theory of identity is a theory about the creation, provenance, comparison and transformation of identifiers. Our analysis is very general and, at the same time, sufficiently precise to be articulated in terms of the formal methods of algebra and logic. We wish to develop a theory that is not about people and objects but about the data that represents their behaviour and identity.

## 2. Life Devolved to Software

Identity is a part of almost every meaningful interaction among individuals. It is so deeply embedded in our daily interactions that we hardly give it much attention (Harper, 2006). The role of identity in surveillance theories varies. Traditionally, the nature of identity is an unproblematic issue. In each context, identity is assumed to be or at least, treated as something stable or unambiguous. Moreover, the range of contexts was limited and tools to resolve questions of identity were effective. Nowadays, identity is extended from the embodied physical individual or object, to their representations in terms of abstract data in multiple databases. Technological development has accelerated the demand to identify product, including people and animals, e.g., the identification of farm animals in the UK (UK Government, 2013). These demands have changed our appetite for data from a vice to a habit, and not merely in a great plurality of identities in multiple databases.

Indeed, as increasing elements of our identities are devolved to technological systems driven by software, the relationships between identity and surveillance become technically and socially intricate. In sociological terms, these technological systems are known as abstract systems. Examples include banking, transportations, e-commerce, mobile telecommunications… and last but not least, social media sites.

Living among these systems, surveillance practices involve the creation of new identities that represent the individual to which they are attached. These 'data doubles', 'data images', 'digital personae' or 'additional selves' are created through surveillance processes (Lyon, 1994; Haggerty & Ericson, 2000; Clark 1994, respectively). In short, an individual's data doubles are various concatenations of personal data that represent the individual in different systems belonging to different organisations. Access to these systems requires an entry in an appropriate database and the presentation of the correct data as a key. In these processes, the data double is the individual's identity for the purpose of that system and that organisation: identity is reduced to codes, passwords or signatures (cf., Deleuzian, 2002).[1] In this way, the practices of identity management originate in online environments have already entered the physical world (O'Hare & Stevens, 2006).

Indeed, surveillance is about data and essential to contemporary surveillance practices are software technologies and hardware devices that collect, store and process data. Currently, the growth and effectiveness of surveillance are made possible by all sorts of new technologies, especially software technologies. They are increasingly becoming a natural component of our everyday life as various forms of surveillance practices are routinely built into our physical and virtual environments. Surveillance is a process of data gathering that involves the systematic observation of behaviours and individuals, and the identification of the ones that are deemed to have specific attributes (see: Figure 1).

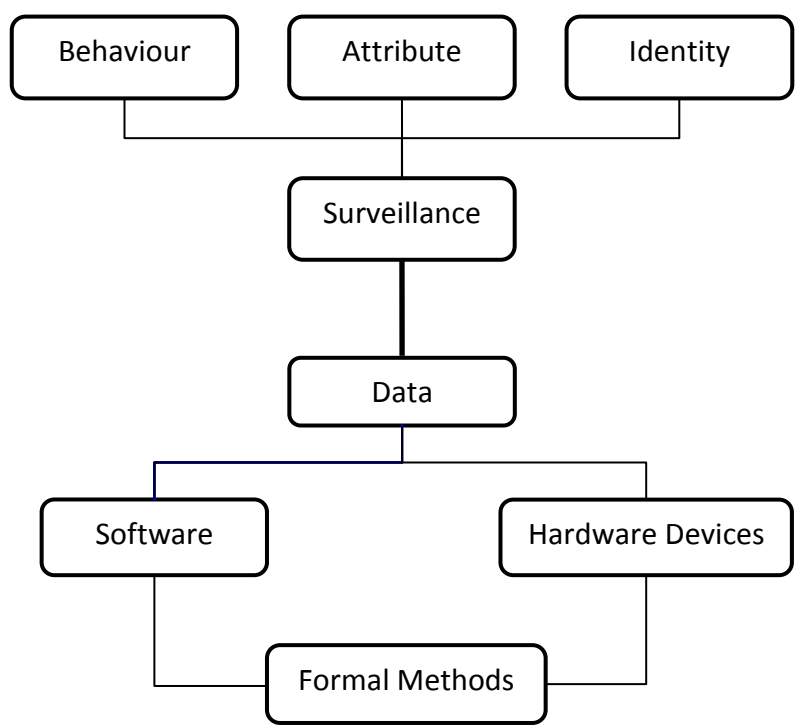

Figure 1: The Conceptual Framework

---

[1] Data doubles are vulnerable to alteration, addition, merging and loss as they travel. The ongoing life of the data doubles depends upon complex information infrastructures.

Many systems that shape and hold together contemporary life are made with software. These technologies have surveillance capabilities, which may be unintended rather than intended. Actually, all software technologies that we use in everyday life naturally collect, store and process data *about their own operation* and, therefore, can be the basis of surveillance technologies. It is widely recognised that computers and software are responsible for the staggering growth of large data sets. Indeed, software is created for users to process data. However, since the earliest computers, designers of software have created tools to monitor the operation and performance of their systems. The tools are necessary to maintain and improve their software. Operating systems that manage computers and their peripherals need diagnostic programs for system administrators to detect and correct faults, install special software, remove malware, and rescue lost data and connections. At low levels of abstraction, the software tools close to machine architectures are used in computer forensics (e.g., for reconstructing memory). The design of diagnostic programs that monitor and analyse systems is natural in software engineering. Thus, where there is software, there are likely to be programs collecting and recording data on its operation and, therefore, *on the activities of its users*. Nowadays, various kinds of software are embedded in many domestic objects from cars to washing machines – called embedded systems. An embedded system is a computer system with a dedicated function within a larger physical system. Typically, it controls a mechanical or electrical system, collecting data and computing via sensors, actuators and processors. Dodge and Kitchin (2009, 2011) offer a taxonomy of some objects employing software. They coin the term *logject* for an object that monitors and records its own use in some ways. The records can be stored and transmitted and so analysed. The smart phone is a primary example: it manages itself, adapts to the environment and is highly programmable by millions of apps.

Examples of software systems intended for surveillance are plentiful. During the past two decades, there has been a significant and continuous growth in the use of surveillance and identity technologies. Actions and events are addressed by Closed-Circuit Television (CCTV), Automatic Number Plate Recognition (ANPR) systems, the Global Positioning System (GPS) and electronic tagging. Personal identities are managed with Radio Frequency Identification (RFID) cards, PIN numbers, coded entry systems, biometric recognition systems and DNA profiling. These forms of surveillance and identity technologies enable the recording, storing and processing of data in digital forms. In turn, the wide spread adoption and combination of such technologies further stimulates the rise of surveillance, in both physical and virtual surroundings, turning the concept of global surveillance into a plausible prospect.

Besides these technologies deliberately created for surveillance purposes, surveillance is often intentionally included in various everyday technologies, such as couriers who deliver packages, inspectors who collect data and operators in call centres. Individuals who work with ICT systems have long been aware that the systems that they use in their work are also intended to be a source of data about their performance – complete recordings of the transactions are made.

In social media sites, such as Facebook, Twitter, Instagram and YouTube, various forms of technologies are used to gather personal data. For example, geotagging is one of the latest forms of tagging that allows real-time surveillance. Often, geotagging is used in (i) searching social media postings on sites by location and finding individuals; and then (ii) exploring their profiles to find out their private information (e.g., home address and phone number) (Smyser and Holt, 2012). In the Android world, an app called eBlaster Mobile can be used to watch over phone usage of children or employees by (i) monitoring text messaging, voice, and Web surfing activity on the Android device; and (ii) logging the physical location of the smartphone (Bradley, 2011).

In more extreme cases, social rules and norms can be approximated by algorithmic formulae to search for deviance automatically, and even render deviance from the roles and norms impossible. For example, various cybercommunities are places where 'dataveillance' (van der Ploeg, 2003: 71) is endemic – every word typed and every movement made can be observed, recorded, stored in digital files, and replayed and examined in the future. Actually, in theory, it is perfectly possible to turn the world of the Internet into a digital panopticon, where surveillance can reach perfection, at least in principle (Wang et al., 2011).

Identity is a core part of existing and developing surveillance practices. The Identity Card Act 2006 is a very good example. The core of this act is the National Identity Register (NIR) of which identity cards are a physical manifestation. The Act's definition of 'identity' refers to 'full name, other names by which an individual might previously have been known, gender, date and place of birth and external characteristics of his that are capable of being used for identifying him' (UK Government, 2006: 2). It introduces a major restructuring of the way identification functions in the UK – identity becomes associated with a singular centralised authoritative documentary source.

Unlike identities in the physical world, data doubles have a much greater mobility than their physical counterparts, especially in terms of reproduction and transmission. Owing to the creation of a polymedia[2] environment brought about by the rise of communications technologies over a range of platforms, we are able to instruct our identities to perform different functions in a digital networked world. The increased use of mobile technologies (smart-phones and tablets) and various apps associated with these technologies have catalysed the explosion of the hyper-connectivity, increasing social plurality, and blurring of public and private identities. The data doubles are creations of information, and are constantly updated due to the information flowing from the individuals (Lyon, 2007a). In extreme cases, they can experience a life of their own and can have significant impact on decision making. Various forms of identities, physical and digital, have the potential to have very real influence on the individual's life chances and opportunities. The core articulation of the problem of surveillance is that identity is problematised. Thus, in the context of a general definition of surveillance, identity must be checked and secured.

## 3. A General Definition of Surveillance

A surveillance system observes the behaviour of people and objects in space and time; classifies behaviours into attributes; and identifies people and objects with some of those attributes. With a wide range of contexts and specific examples in mind, we begin to analyse four concepts that capture the structure of surveillance systems.

***Definition.*** Abstractly, we say that a *surveillance system* consists of the following components and methods:

1. ***Entity.*** Entities that are people or objects that possess behaviour in space and time;
2. ***Observable behaviour.*** Methods for observing and recording behaviours;

---

[2] A condition where individuals are free to choose between a number of equally available forms of media, such as Facebook, Twitter, MySpace … (Madianou & Miller, 2012).

3. ***Attribute.*** Methods for defining and recognising attributes of behaviours, based on rules, norms, practices, and other observable properties; and
4. ***Identity.*** Methods for generating data that identify entities which exhibit the attributes and locate them in space and time.

Commonly, we expect the attributes to be *deviations* from sets of rules, norms, practices, etc. However, the definition does not imply deviance: the definition does require precise formulations of attributes. The data that identifies entities include numbers, texts, sounds and images.

First, we look at particular examples of surveillance systems to see these concepts *in situ*: (i) cars; (ii) smartphones; and (iii) Twitter.

***Example 1: Cars.*** ANPR is a technology that observes cars and recognises number plates, possibly using infra-red so as to function day and night. Common applications are checking on vehicle speed, managing car parking and collecting tolls. The technology was functioning in the late 1970s; today, ANPR can be found in thousands of fixed surveillance systems owned by both public and private organisations.[3] We describe some British ANPR applications to test our abstract definition.

The entities in such surveillance systems are cars at a particular location and time; they may be in transit (speed check), or entering or leaving a location (car park, city zone).The method the system uses for observing the cars is a camera that creates an image that may be communicated and stored. This image is processed by software that will recognise a behavioural attribute (e.g., breaking a speed limit) and, in particular, performs optical character recognition to establish the registration mark of a car. The registration mark is an alpha-numeric name that identifies a vehicle uniquely. On communicating this registration mark, the identity of the entity is established. For example, the output of such surveillance systems is the identity of a vehicle travelling too fast, or arriving or leaving a particular location. For example, a surveillance system for car parking based on an ANPR consists of:
*Entity*: Cars

---

[3] The number of fixed speed camera sites in England increased to 2,331 in 2012 (Massey, 2012).

*Observable Behaviour*: Arrival and departure at location

*Attributes*: Duration of stay above a particular limit

*Identity*: Registration marks

Actually, it is important to note that such surveillance systems deliver identities of vehicles rather than people. The process of observing the behaviours of drivers is actually a process of matching people to data. Following the ANPR stages described above, the registration mark is communicated to a database relevant to the application. For example, the database may be used to check an attribute, such as a payment (tax, charge or toll) having been made for that registration mark. If a deviant attribute is detected then, in order to find the driver, the keeper of the vehicle must be located and contacted. In the UK, the operator of the surveillance system communicates the registration mark to the Driver and Vehicle Licensing Agency (DVLA) to determine the name and address of the keeper. The output of this process is the identity of the keeper. Finding the driver may require further action. Note that in this interpretation there is a *transformation of identity data*; identity changes from the registration mark to the name and address of the keeper. Thus, coupled to the surveillance system are independent systems processing identities, in order to locate a person.

***Example 2: Smartphones.*** A smartphone is a mobile phone built on an operating system (e.g., Android, iOS and BlackBerry 10). A modern smartphone includes various types of advanced software that could be used for surveillance purposes, including a compact digital camera; a pocket video camera; a Global Positioning System (GPS); an accelerometer; high speed data access provided by Wi-Fi and mobile broadband; and last but by no means least, all sorts and kinds of apps. These technical features (i) allow smartphones to be used by their owners as spies in their pockets to surveil; and, conversely, at the same time, (ii) these individuals could be surveilled upon when they are carrying their smartphones with them. Common applications of smartphones are to catch offenders (e.g., intelligence and evidence gathering); to identify risks to public safety; and to gather otherwise hard-to-get data (peer-to-peer mutual monitoring) (O'Keeffe, 2011).

In the second case, consider a user of a smartphone who is surveilled upon. The technologies in the smartphone (e.g., GPS and accelerometer) make it possible for the movement of the smartphone to be tracked to a high degree of accuracy. Technically, a constant stream of data is being created and communicated by the smartphone to the service provider. All sorts of properties of the user can be deduced from these data streams, such as conversing, exact location (GPS) and even running/walking/driving (accelerometer). The entities are smartphones and the observable behaviour is the activity and location of the smartphones. The attributes are what interests the surveillance agent, e.g., an activity or a location; and the identities are numbers of the smartphones.

It is easy to describe a few smartphone applications in terms of our abstract definition. Consider first the case of using a smartphone to surveil. For example, imagine someone recording an activity and streaming the video live to a website (e.g., YouTube) – remarkably but not uncommonly, an Internet TV channel with live broadcasting can be created from a smartphone in one's pocket. More simply, the smartphone may capture an isolated activity that is at risk to public order and safety. The entities may be people or objects and the observable behaviour is simply their activity. The attributes are what define risk and the identities are represented by images and sound. This rather direct form of surveillance delivers data, which is the start of independent processes for identifying people.

Moreover, these properties, e.g., location and activity, of the user's actions can be combined with other data in surveillance to provide more detail about him/her (e.g., mood) and predict his/her action. Given that smartphones are ubiquitous and serve personal needs, they have tremendous potential in all aspects of surveillance, especially in terms of identifying individuals. Actually, a network of smartphones could provide the police with thousands of eyes and ears on the streets at any one time, and thus, could create a big society where criminals will live in fear of the people and where there is nowhere for them to hide (May, 2010). Conversely, the network would generate an increasingly large amount of data and produce more data than necessary for the originally stated purpose, and in turn, would lead to complicated social questions concerned data storage, access and use.[4]

Again, such surveillance systems that employ the use of smartphones as spies deliver data about devices rather than people. In this section, we have occupied ourselves with surveillance "in relation to non-human phenomena that have only a secondary relevance to personal details" (Lyon, 2007a). Next, we turn our attention to surveillance in relation to the actual person.

---

[4]For an example, Apple's iPhone and iPad, Google's Android and Microsoft's Windows Phone 7 operating systems are designed to collect, store and transmit data on users' physical locations to central databases without their consent (Burghardt, 2011).

## 4. A General Definition of Identity

The examples of the last section illustrate the four conceptual components of our general definition of surveillance systems. They also reveal the necessity of additional systems to process their output data. Commonly, the output of a surveillance system is data about the identities of devices and machines rather than their users. At this point, a few questions arise:

- What is the nature of the data that is supposed to identify the entities under surveillance?
- How is this data transformed to yield information relevant to the context?
- How do this transformations lead to the identity of people?

To explore these questions, we will examine the nature of identity with respect to surveillance largely by means of examples: (i) cars; (ii) communications; and (iii) customer accounts. Currently, individuals tend to have several over-lapping identities, some of which could be attributed to developments in technologies, especially these associated with the Internet, such as social networking sites (e.g., Facebook and Twitter) and cybercommunities (e.g., Second Life and Cyberworlds). A few years ago, McGuire coined the term 'distributed body' and used it to explain the relationship between the body in the physical world and 'some new category (virtual or otherwise)' (2007: 82). These new categories simply extend the physical body's ability to interact by distributing it across more spheres, including bank accounts, my space pages, blogs, mobile phone numbers and game personas (ibid.).

Actually, much of our social interaction is carried out by various forms of abstract technological systems rather than direct face-to-face interactions. Whilst interacting with each of these systems, an individual needs to give over some of his/her identity, in terms of identifiers, to distinguish himself/herself from other users. Thus, rather than having a single and holistic identity, individuals now have several separated and overlapping identities. The multiplicity of identity, especially the extension of identity from the physical to the virtual world has brought us to a new era, wherein the nature of identity needs to be problematised.

The rise of virtual identities in various forms of abstract systems, professionally and personally has generated new forms of deviant behaviours. This is more so in the virtual world since it is created through the notion of anonymity, as well as social and technical appetite for multiple identities.

## 4.1. On Identifiers

Once verified, identities provide access to resources and therefore, become a thing or object of some kind. This process gives identity a value that can be traded, sold or stolen. Identifiers become assets. Basically, they are snapshots of an individual whose purpose is to allow access to resources, typically through information systems. Be it physical or virtual, an identity is presented by a data type. *Basic personal identifiers* are those upon which we rely to distinguish a unique human being and guarantee their identity, albeit in some context with its own level of rigor. Essential to establishing identity is the collection, storage and processing of data. Indeed, identity is almost purely a matter of data. People and objects are named, labelled or otherwise denoted by data relevant to some context. The data in question captures some relevant aspects of a person or an object. Different identities are managed by different kinds of *identity management systems*. We will look at some examples of these systems prior to providing a working definition. Our examples involve birth certificate, health records, driving license and National Insurance (NI) number, and demonstrate that identifiers are composite objects, in the sense that they are built from other identifiers.

Thus, the purpose of an identifier is to establish when entities are the same or not in the surveillance context. Identifiers need not reflect any aspect of the entity or have any meaning at all. In our conception of surveillance, entities are observed and identified. This means that necessarily, surveillance systems must have methods to define the identity of entities.

*Definition.* An *identifier* for an entity is a name that is associated with the entity. By a name we commonly mean data made from symbols. In terms of symbols, usually, numbers are added to identifiers in order to make an identifier unique.

The relationship between entities and identifiers can be complicated. Logically, four situations can arise:
1. *Many – One Associations*. Different identifiers can be assigned to the same entity.

2. ***One – One Associations***. Different identifiers are assigned to different entities.
3. ***One – Many Associations***. An identifier can be assigned to more than one entity but an entity has only one identifier.
4. ***Many – Many Associations***. An identifier is assigned to more than one entity and, vice versa, an entity can be assigned more than one identifier.

We have defined surveillance to be a process that observes the behaviour of entities and reports entities with interesting behaviour. Actually, surveillance reports identifiers which may narrow the search for entities but need not pin down the particular entity of interest.

***Search Principle:*** *If an association is many-one then given an identifier, we can find or search for a set of entities with that identifier.*

One-to-one associations are important because:

***Uniqueness Principle:*** *If an association is one-one then given an identifier, we can find the unique entity with that identifier.*

The following point may be obvious but it is certainly profoundly important:

***Enumeration Principle:*** *The addition of a number to an identifier of an entity can turn any many-one association into a one-one association.*

We will illustrate these as follows:

***Example 1: Cars.*** This example illustrates one-one and many-one associations. Each car is assigned a registration mark, commonly known as registration number. The current system of UK was introduced on 1$^{st}$ September 2001. The association of registration marks to cars is one-one. A car has one and only one registered keeper. Thus, the association of a registration marks to a keeper is unique. However, a person can be a registered keeper of as many cars as he/she wants. Thus, the association of registration marks to keepers is many-one.

The registration document (V5) for a car identifies the car and its keeper. However, it is not proof of ownership. The registered keeper is the person who is legally responsible for the car and need not to be the owner of the vehicle. Many people have insurance policies that enable them to drive any car with the owner's permission. Thus, the driver of a car on a particular occasion may be only loosely connected to the keeper. The association of cars to drivers is one-many. In terms of formal documents (containing several identifiers), the association between registration marks and drivers is complicated and probably incomplete.

***Example 2: Communications.*** This example demonstrates both many-one and many-many associations. When connecting a computer to the Internet, a number is needed called an IP address (32 bits under IP Protocol 4) that uniquely identifies the machine in the network. In some computer networks, such as networks local to an organisation or company, there is an IP address for the machine that does not change; these are called static IP addresses. The association of computers to IP addresses is one-one. More commonly, at home IP addresses are generated by the Internet Service Provider in response to a customer's need for Internet access. Thus, overtime IP addresses can change and the association of IP addresses to a particular computer is many-one. Developing this example, if more than one computer is accessing the Internet at the same time in a period, from the same service then the association between IP addresses and computers is many-many. The changing status seems to be natural in time-dependent associations of identifiers.

***Example 3: Customer Account.*** Consider a client's account with some service company, such as a bank, insurance company or shop. Commonly, such an account has the following structure (see: Figure 2). The user name and password act as a key simply to gain access to the account. The account details establish basic information such as: name, address, services provided, etc. The account history not only records the past transactions but allows all sorts of new transactions, queries, etc. to be performed. It is the account history that is clearly subjected to tests that ensure, for example, terms and conditions are met by the client or that no unusual pattern of transactions has been carried out. If each customer account for a provider belongs to one and only one person then the association of customer accounts to people is one-one. If a customer account can be made available to more than one person – such as a joint bank account – then the association of customer accounts to people is one-many. It is common that whilst accounts may be one-many the systems of identifiers for the accounts – involving numbers – are one-one.

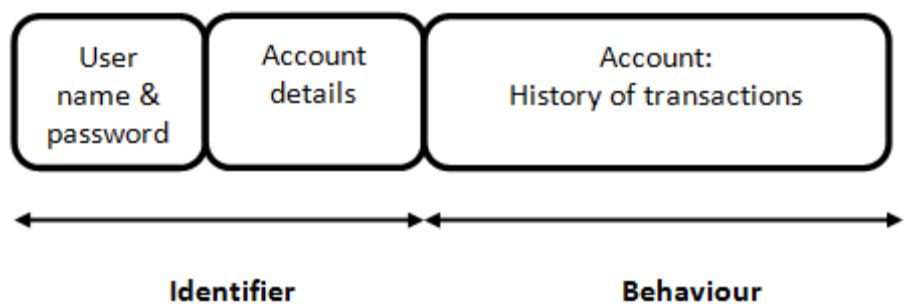

Figure 2: A Typical Customer Account

### 4.2. Provenance of identifiers

Creating identifiers is an everyday occurrence; we open accounts, register for services, apply for permissions, buy products, etc. For many of these actions, we rely on a handful of pre-existing identifiers. To open a bank account in the UK, we give a proof of our identity and our current address, e.g., using a passport and a recent utility bill. To buy a product, an address and a credit card account number are usually sufficient for the vendor: notice the dependency on the bank identifier. At face value, the quality of a bank identifier is guaranteed by the databases of the state (passport) and, say, an energy provider (utility bill). The passport provides a high quality identifier based on a birth certificate, a photograph and possibly other biometric data. Example by example, illustrates the general point that:

***Principle.*** *The creation of new identifiers depends on pre-existing identifiers.*

The dependability of one identifier upon another may be illustrated in an *identity tree* (see: Figure 3). The identifiers that appear in the nodes of the tree can create quite complicated dependency networks of identifiers.

The quality of an identifier is essentially a matter of its reliability, which in turn depends on its provenance, i.e., the process involved in establishing the identifier. In the case of people, a passport is a standard example of a high quality identifier with a rigorous provenance. Since identifiers are often built from other identifiers, of central importance is the process of comparing identifiers and reducing one type of identifier to another. Recognising a number plate of a car behaving badly can lead to a letter arriving at the address of the keeper and involves the transformation of a number of high quality identifiers (e.g., registration mark, keeper's name, address and driving history).

In Figure 3 below, establishing the identifier ID1 involves providing evidence in the form of other identifiers: ID2-ID6. Thus, the validity of ID1 depends upon, or is reduced to, the validities of ID2-ID6. Some of these identifiers have a special status, in that they are designed to reliably denote an individual. In the example, these personal identifiers are guaranteed by the state (ID4) and biometric data (ID3); in the latter case, ID6 is used to allow a passport to be issued by post, without face-to-face interaction.

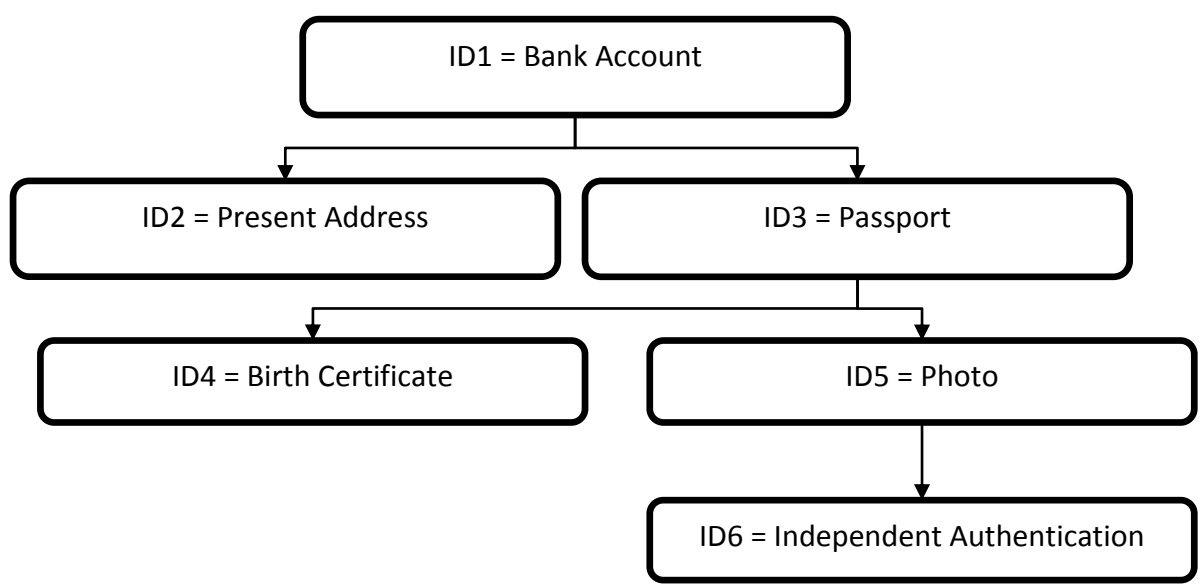

Figure 3: Identity Tree

This example illustrates a general fact about identifiers:

***Principle***: *To establish the provenance of an identifier is to follow a set of paths belongs to a tree of identifiers that represents the construction of the identifier.*

All of these observations and ideas can be formalised to make a precise and general mathematical framework for analysing identifiers.

5. **Personal Identity**

14-08-2014

Of greatest interest are surveillance systems in which the entities are people, for example, birth certificate and photo (see: Figure 3). We consider some examples of assigning data to individuals. A fundamental problem is how identifiers can actually identify a specific individual.

***Example 1: Biometrics.*** Biometrics refers to the identification of humans by their characteristics or traits. Biometric identifiers are the distinctive, measurable characteristics used to label and describe individuals. Biometric identifiers are often categorised as physiological versus behavioural characteristics. Physiological characteristics are related to the body, including fingerprint, face recognition, DNA, palm print, hand geometry, iris recognition, retina and odour/scent. Behavioural characteristics are related to the behaviour of a person, including typing rhythm, gait and voice. Operationally, as evidence in practice, the associations of biometrics to people are assumed to be one-one.

The operational tests used to measure biometrics, such as DNA, finger print and iris, are of course, approximate. Thus, it is a matter of *high probability* that data presented manifests a one-one identity association. Current studies suggest that increasingly accurate measurements can reveal differences in DNA between, even, twins. Lately, research demonstrates that although identical twins share very similar DNA, they are not identical (O'Connor, 2008). Recently, when identical twins are identified by DNA evidence as suspects in a series of rapes in Marseille, France, officials are likely to pay about $1.3million to compare billions pairs of nucleotides that make up DNA rather than compare 400 base pairs in a normal analysis (Dicker, 2013).

***Example 2: Citizenship.*** Each individual has or can have a unique passport, driving licence, National Health Service (NHS) and National Insurance (NI) number. In the new style red passport book, the passport number must be nine characters and all characters must be numeric. Each driving licence in England, Scotland and Wales is made up of 18 alpha-numerics, which codes a great deal of information. Everyone registered with the NHS in England and Wales has his/her unique number, which is linked to his/her health record. Each NHS number is made up of 10 alpha-numerics in a 3-3-4 format. In the UK, everyone gets a NI number just before he/she turns 16. An individual's NI number makes sure his/her NI contributions and taxes are only recorded against her/her name. The format of the number is two prefix letters, six digits and one suffix letter. In these cases, usually numbers are added to identifiers in order to make each of these associations one-one.

***Example 3: Social Media.*** Each individual has a unique user name and password combination to log into social media sites, such as Facebook, Twitter and MySpace. For example, to sign up for a new account on Facebook, an individual needs to enter his/her name, birthday, gender and email address into an online form on [www.facebook.com](www.facebook.com) prior to picking a password. After he/she completes the sign up form, an email would be sent to the email address provided. The sign up process is completed by clicking the confirmation link embedded in the email. The next stage to create a basic profile to highlight some key characteristics of a person, such as basic information (birthday, relationship status, religious views…), work and education, relationships and family, and contact information.

The basic structure of a user name and password combination reminds us of our previous discussion on customer account (see: Figure 4). Actually, a person's social media account is essentially a client account with some service provider, e.g., Facebook. However, it is not the same as a customer account. In a social media account, a person's 'User name & password' and 'Basic personal profile' form his/her identifier, instead of 'User name & password' and 'Account details' in a typical customer account. Communication and interaction with other individuals is key in having a social media account, thus 'interaction history' forms the last part of a typical social media account (recalling the third part of a typical customer account: 'account history'). In the case of Facebook, a person's 'interaction history' covers all his/her activities on Facebook, such as updating his/her status, commenting on another's post, status or photo, sending another user a private message… The 'interaction history' is a part of the person's personhood – the physical body's ability to interact is extended into the virtual sphere of Facebook (McGuire, 2007).

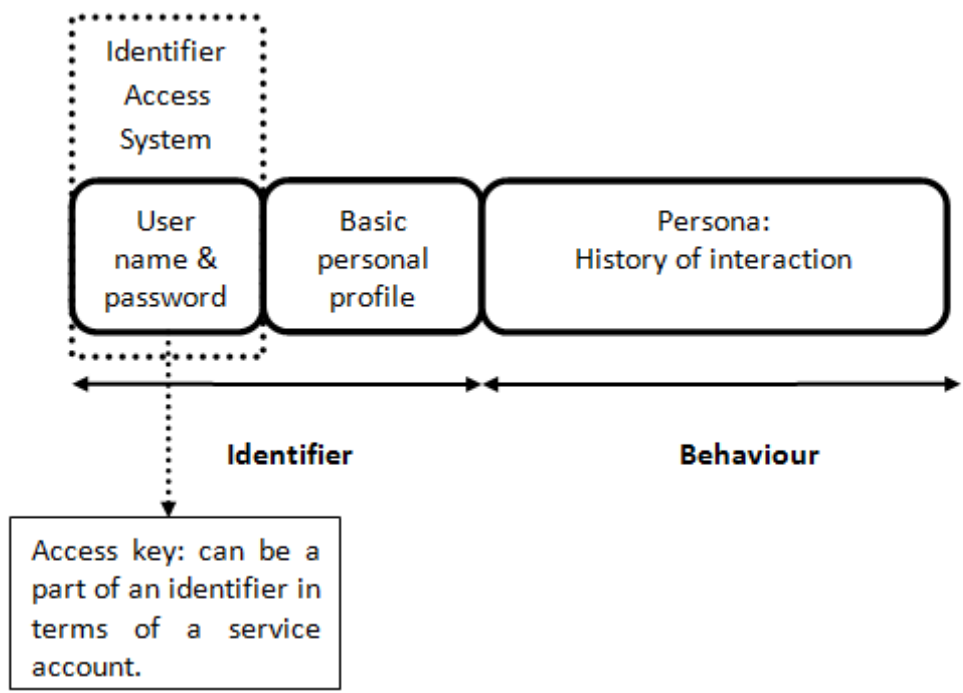

Figure 4: A Typical Social Media Account

## 6. Surveillance revisited

With the concept of identity clarified we can revise the provisional definition of surveillance. First, let us define an identity management system.

***Definition.*** An *identity management system* for a set of entities is a system with the following two properties:

(i)   *Generation*: the system can create and delete identifiers for entities; and

(ii)  *Entity Authentication*: the system can, given an entity and identifier, decide whether or not the identifier is associated with the entity.

Another formulation of authentication, which focuses *exclusively* on the identifiers, is the following:

(iii) *Identity Authentication*: the system can, given two identifiers, decide whether or not they are associated with the same entity.[5]

---

[5] Property (ii) implies property (iii).

Our discussion of this is independent of application, since the identifiers can often be used to gain access to resources. We are only dealing with access not authorisation.

***Definition.*** A *surveillance system* is a method of observing the behaviour of entities in a certain context. The context is specified by:

(i)     a collection of attributes of the behaviours;

(ii)    an identity management system that provides identifiers for the entities.

Surveillance is then a process that, on recognising that the behaviour of an entity has an attribute, returns some identifier for that entity.

The concept of social sorting can be formulated in two ways. At first sight, we might say that it is simply the process of putting entities into groups or categories defined by some properties of their behaviour that they have in common. This means that the categories are defined by attributes of behaviour that are specific to the context of surveillance system (rather than any intrinsic nature of the entity). The intention that the different categories of entities are to be treated differently is not part of the abstract definition.

The social sorting is output of the surveillance. However, in our conception of surveillance, the process outputs not entities but identifiers for entities. Thus, we propose the following:

***Definition***. A surveillance system is called a *social sorting* if it is a method of classifying the behaviour of entities in a certain context. The context is specified by

(i)     a collection of attributes of the behaviours;

(ii)    an identity management system that provides identifiers for the entities.

Then the surveillance system provides a process that builds a collection of categories of identifiers. On recognising that the behaviour of an entity has a particular set of attributes the system places *some* identifier for that entity in the category defined by those attributes.

The relationship between identifiers and entities can be complicated and the variety of classifications of the identifiers for entities even more so. The definition above is deliberately general. It permits the surveillance system to generate the classification by combining attributes from a list in different ways. It does not require all the possible identifiers of an entity to be in the classification. It allows the different categories to overlap and even allows one category to be a subcategory of another.

## 7. Conclusion

Identifiers are simply data and belong at the heart of any analysis of surveillance. Our research on surveillance, with its emphasis on identity, is intended to develop an abstract framework that is both a rigorous conceptual analysis of surveillance and a tool for answering questions about applications. We have isolated a general structure or architecture that can be found in a large number of apparently disparate surveillance situations – certainly including three general typologies of surveillance: controlling, sorting and monitoring (Lyon, 2007a). A primary feature of our framework is the combination of systems for observing and categorising behaviour, and managing identity. In particular, the abstract notion of identifier enables us to make explicit the complexities of establishing personal identity. Identifiers seem always to be dependent on other identifiers: verifying identity involves following paths through a network of inter-related identifiers.

There are several more basic topics that need to be analysed and added to this framework. Although we have accommodated social sorting in our framework, its abstract analysis is clearly a necessary and complex next step. Another example is to formulate general concepts and principles for comparing identity management systems: in particular to structure the process of reducing or translating one identity management system into another.

Common to all surveillance technologies are software systems. This is because the collection, storage and processing of data is the 'raison d'être' of software. The study of data is the basis of the science of surveillance. In particular, in surveillance, we must study:

(i)     the representation of various forms of data (visual, sonic, audio and textual);
(ii)    the construction, provenance and validity of data;
(iii)   the transformation and aggregation of data;
(iv)    access control, communication and privacy of data; and, of course; and
(v)     the huge variety of computation of these data.

The latter point is important because most the world's data is currently or destined to be computed, rather than collected. By computed we mean generated via mathematical models; and by collected we mean gathered by direct forms of observation, survey and measurement.

The rise of data science can be attributed to the success of using large volumes of collected data to created large volumes of computed data to drive applications.

In extremis, let us observe that whenever a social science topic – in this case surveillance – is closely associated with technology – especially with technological tools that collect and process data effectively – then the specification of the software tools, i.e., what they do for users, can be analysed in much the same way as we have done here. Thus, the sociological notions that motivate and shape, and must be represented in the specification of software must be as precise as possible. The act of analysing general notions abstractly is commonplace in areas of physical sciences, philosophy and linguistics but seems to be rare in the social sciences. Keeping in mind some contemporary examples of surveillance technologies, we have modelled the main abstract ideas involved in surveillance. At a later stage, we foresee the use of formal mathematical methods to analyse much more precisely and, in great generality, the sort of fundamental concepts, structures and processes that we have isolated in this article.

14-08-2014

**Reference**

Ball, K., Haggerty, K. & Lyon, D. (ed.) (2012). *The Routledge Handbook of Surveillance Studies*. London: Routledge.

Burghardt, T. (2011). '*Smartphones: The Tracking and Surveillance of Millions of Americans*'. World Truth Today, 3rd, May.

Clark. R. (1994). The Digital Persona and its application to data surveillance. *The Information Society*, *10*(2): 77-92.

Cole, S (2001). *Suspect Identities: A History of Fingerprinting and Criminal Identification*. Cambridge, Mass & London: Harvard University Press.

Dicker, R. (2013). Identical Twins' DNA Evidence in Serial Rape Case Confounds Police in Marseille, France. *The Huffington Post*, 11th, February.

Dodge M. & Kitchin R. (2009). Software, objects, and home space. *Environment and Planning A, 41*(6): 1344 – 1365.

Dodge M. & Kitchin R. (2011). *Code/Space*. MIT Press.

Foresight (2013). *Future Identities Changing identities in the UK: the next 10 years (Final Project Report)*. Government Office for Science.

Introna, L. & Wood, D. (2004). Picturing Algorithmic Surveillance: The Politics of Facial Recognition Systems. *Surveillance & Society, 2*(2/3): 177-198.

Haggerty, K. & Ericson, R. (2000). The Surveillance Assemblage. *British Journal of Sociology, 51*(4): 605-662.

Harper, J. (2006). *Identity Crisis: How identification is overused and misunderstood*. Washington: Cato Institute.

Lyon, D. (1994). *The Electronic Eye: The Rise of Surveillance Society*. Cambridge: Policy.

Lyon, D. (2001). Under my skin: from identification papers to body surveillance. In Caplan J (ed.) *Documenting Individual Identity: the Development of State Practices in the Modern World*. Princeton Univ. Press.

Lyon, D. (ed.) (2003). *Surveillance as Social Sorting: Privacy, Risk and Digital Discrimination*. London and New York: Routledge.

Lyon, D. (2007a). *Surveillance Studies: An Overview*. Malden, MA: Polity Press.

Lyon, D. (2007b). *Surveillance as Social Sorting*. (2007, Sept 21). Retrieved from http://www.youtube.com/watch?v=xtAa-f-1rTg.

May, T. (2010). *Theresa May's speech to the Police Federation*. Home Office, 19 May 2010.

McGuire, M. (2007). *Hypercrime – The New Geometry of Harm*. Routledge-Cavendish.

Madianou, M. & Miller, D. (2012). Migration and New Media: Transnational Families and Polymedia.

Massey, R. (2012). So much for the promise to cut speed cameras… now there are even more! *MailOnline*, 10th, July.

O'Connor, A. (2008). The Claim: Identical Twins Have Identical DNA. *The New York Times*, 11th, March.

O'Hara, K. & Stevens, D. (2006). *Inequality.com: Power, Poverty and the Digital Divide*. Oxford: Oneworld.

O'Keeffe, C. (2011). 'Smartphone surveillance makes spies of all citizens'. *Irish Examiner.com, 19 September 2011*.

Smyser, K. & Holt, S. (2012). Geotagging Allows for Real-Time Surveillance. *5 NBC Chicago*, 20th, November.

Torpey, J. (2000). *The Invention of the Passport: Surveillance, Citizenship and State*. Cambridge: Cambridge University Press.

UK Government (2006). *The Identity Card Act 2006: Elizabeth II. Chapter 11*, London: The Stationary Office.

UK Government (2013). Animal Identification, Movement and Tracing Regulations, London: The Department for Environment, Food & Rural Affairs.

van der Ploeg, I. (2003). Biometrics and the body as information: narrative issues of the socio-technical coding of the body. In Lyon D (ed.) *Surveillance as Social Sorting: Privacy, Risk and Digital Discrimination*. London: Routledge.

Wallace, H. (2006). 'The UK National DNA Database – Balancing crime detection, human rights and privacy. *Science & Society, EMBO reports 7*: S26-S30.

Wang, V., Haines, K. & Tucker, J.V., (2011). Deviance and Control in Communities with Perfect Surveillance – the Case of Second Life. *Surveillance and Society 9*(1/2): 31-46.